\documentclass[preprint,12pt]{elsarticle}

\usepackage{amssymb}
\usepackage{geometry}

\journal{Computational Materials Science}

\begin{document}

\begin{frontmatter}



\title{Convolutional neural networks for atomistic systems}

\author{Kevin Ryczko}
\address{Department of Physics, University of Ottawa}
\ead{kryczko@uottawa.ca}
\author{Kyle Mills}
\address{Department of Physics, University of Ontario Institute of Technology}
\author{Iryna Luchak}
\address{Department of Electrical \& Computer Engineering, University of British Columbia}
\author{Christa Homenick}
\address{National Research Council of Canada}
\author{Isaac Tamblyn}
\ead{isaac.tamblyn@nrc.ca}
\address{Department of Physics, University of Ontario Institute of Technology}
\address{Department of Physics, University of Ottawa}
\address{National Research Council of Canada}

\begin{abstract}
We introduce a new method, called CNNAS (convolutional neural networks for atomistic systems), for calculating the total energy of atomic systems which rivals the computational cost of empirical potentials while maintaining the accuracy of \emph{ab initio} calculations. This method uses deep convolutional neural networks (CNNs), where the input to these networks are simple representations of the atomic structure. We use this approach to predict energies obtained using density functional theory (DFT) for 2D hexagonal lattices of various types.  Using a dataset consisting of graphene, hexagonal boron nitride (hBN), and graphene-hBN heterostructures, with and without defects, we trained a deep CNN that is capable of predicting DFT energies to an extremely high accuracy, with a mean absolute error (MAE) of 0.198 meV / atom (maximum absolute error of 16.1 meV / atom).  To explore our new methodology, we investigate the ability of a deep neural network (DNN) in predicting a Lennard-Jones energy and separation distance for a dataset of dimer molecules in both two and three dimensions. In addition, we systematically investigate the flexibility of the deep learning models by performing interpolation and extrapolation tests.
\end{abstract}


\end{frontmatter}



\section{Introduction}

Solving the electronic structure problem has long been of interest to researchers in materials science, chemistry, and physics.  Even modest systems consisting of a few atoms are impossible to treat exactly and simplifications or approximations must be made to reduce the complexity of the problem.  This could be in the form of the Born-Oppenheimer approximation \cite{Born1927}, frequently used in conjunction with Kohn-Sham DFT \cite{Equations1965}, or the use of phenomenological fits to a set of experimental or theoretical results.  
Although approximate electronic structure methods have the advantage that they preserve the characteristics of the underlying physics (e.g. the wavefunction or ground state charge density is treated as a fundamental object), they are limited in applicability due to unfavourable scaling with system size and computational cost \cite{Ratcliff2016b}.  In the domain of phenomenological fits, force fields informed from high-level theory calculations and experiment have seen success \cite{Mercado2016,Cherukara2016,Kulkarni2015,Jaramillo-Botero2014,VanBeest1990,Ponder2003,Hornak2006} in modelling phenomena that occur on time and length scales beyond the reach of the higher level electronic structure methods.  Force fields even predate the quantum theory itself; the van der Waals equation of state is dependent upon two species-specific fitting parameters argued for based on microscopic atomic interactions \cite{Waals1873}.  Interaction terms evocative of the van der Waals parameters still appear in many modern force-fields \cite{Perez2007,Allinger1989,Allinger1977,Weiner1984}.

Approximations and phenomenological fits are useful in materials discovery and design, where oftentimes a specific property (e.g. band gap, ionization energy, etc.) is desired.  Targeting the search at novel materials with a specific property is a difficult task given the large search space spanned by permutations of atoms.  Thus the ability to make rapid, accurate predictions about prospective materials is invaluable.  Although the term ``machine learning'' has not traditionally been applied to phenomenological fits, the task of reproducing a generalized mapping of input-to-output through a series of observations is the core of supervised machine learning.


Many recent studies have used machine learning in some form to study molecular or condensed matter systems \cite{schutt2017quantum, faber2016machine, chmiela2017machine, Gomez-Bombarelli2016, rupp2015machine, ramakrishnan2015big, huo2017unified, Artrith2016135, artrith2017efficient, LBWB16, BVLT17, rupp2012fast, li2016understanding, bartok2013representing}. 
In particular, G\'omez-Bombarelli et al. \cite{Gomez-Bombarelli2016} used auto encoders to project their training set onto a latent space.  They were able to operate within this latent space, exploring chemical space (by proxy), and found new molecules with desirable properties that were not present in the original data set.  The input to their auto encoder was based off of the simplified molecular input line entry system (SMILES)\cite{weininger1988smiles}, a descriptor consisting of a minimal string of text to describe the three dimensional molecule. Other works using kernel ridge regression (KRR) \cite{chmiela2017machine, faber2016machine, rupp2015machine, ramakrishnan2015big, huo2017unified}, rely on abstract constructions of input feature vectors describing the chemical system. Rupp et al. \cite{rupp2015machine} used principal component analysis (PCA) to obtain a three dimensional atom-centered local coordinate system that was then used as input into KRR. This framework was successful in predicting nuclear magnetic resonance (NMR) chemical shifts, core ionization energies, and atomic forces. Another choice of a feature vector is the one given by Behler and Parrinello \cite{behler2007generalized}. This feature vector was constructed with an artificial neural network (ANN) architecture in mind, and is written in terms of symmetric functions that obey rotational and translational invariances. Preceding this descriptor, Bart{\'o}k \emph{et al.} \cite{bartok2013representing} showed that a new approach, called Smooth Overlap of Atomic Positions (SOAP), eliminates the need for ad hoc descriptions of atomic environments. They showed that by directly defining the similarity between atomic environments, they could still include symmetric and invariant properties, necessary to describe atomic environments. This approach was then successfully applied to fit the potential energy surfaces of different silicon structures, and was also successfully applied in another report \cite{de2016comparing} to traverse through chemical space and make energy predictions for small molecules within chemical accuracy. Additional works include using a Coulomb matrix to make predictions of atomization energies \cite{rupp2012fast}, and understanding of machine learning density functionals \cite{li2016understanding}.

Our alternative approach to overcome the challenge of finding a suitable input feature vector is inspired by the recent successes of applying ``big data'' to grand challenge problems in computer vision and computational games \cite{Silver2016,Mnih2013}. Rather than seeking to simplify, compress, or approximate the interactions within a system, we train a highly flexible, data-driven model on a large number of ``ground truth'' examples.  Here we argue that this brute force approach may offer a more scalable and parallelizable approach to large-scale electronic structure problems than existing methods can offer. In our approach, we use a simple, image-based approximate representation of the electrostatic potential, that preserves spatial structure.  By avoiding the construction of an input feature vector, we sidestep any possibility of introducing biases, and preserve the spatial correlations between atoms in the training examples.

While many machine learning algorithms exist, we have focused on a particular class: deep CNNs. 
CNNs have the ability to ``learn'' non-linear input-to-output mappings without prior formulation of a model or functional form. This is done through the use of convolutional layers. A convolutional layer consists of a set of kernels (matrices) which contain adjustable parameters. When an operation is performed on an image with a convolutional layer, the operation produces a different image for every kernel in that layer. The kernel moves from pixel to pixel in the image, and performs the dot product between the weights and the pixels enclosed by the size of the kernel. The result is a convolved image. While training, the parameters in the kernels are updated so that they have the ability to enhance features in the images necessary for making accurate predictions. Incorporating convolutional operations allows the neural network to exploit the spatial structure naturally present in the input data, and drastically reduces the number of redundant trainable parameters (compared to a traditional ANN).  During the training procedure, the deep neural network automatically ``learns'' by optimizing a set of features necessary to reproduce the desired input-to-output mapping. Recently, Mills \emph{et al.} \cite{mills2017deep} was able to reach chemical accuracy, applying a deep CNN to the one-electron Schr\"odinger equation in two dimensions. The input to the system was solely the external potential. During  the training process, the neural network used the information in a large collection of these potentials to develop a set of features necessary to reproduce the energies.  The features that a deep CNN develops are not directly interpretable, but collectively form a latent space in which the desired mapping can be interpolated.  Deep neural networks excel at interpolation within the latent space, but perform poorly when extrapolating (as we show in Subsection \ref{interp_and_extrap}). One disadvantage of our approach is the discretization of the atomic coordinates on a real space grid. This means that the convolutional kernels that are applied onto the image are also discrete. Additionally, the images that are fed into our network must have the same dimensions as the images the model was trained on. In a recent study \cite{schutt2017moleculenet}, Sch{\"u}tt \emph{et al.} avoided discretization errors by using a continuous convolutional approach. This approach was successful at predicting both energies and forces for small molecules, and avoided constructing atomic fingerprint functions by optimizing the fingerprints in the training phase. Similar to previously mentioned work, information of the local environment (e.g. radial distance of neighbouring atoms) was needed to construct the convolutional kernels. In our approach, the optimal environmental features are calculated during training.

In this Article, we first describe our new method to calculate total energies of atomic systems. This includes how to construct images that are used as input to the CNNs. These images describe the atomic environment, and can be generalized to \emph{any} atomic system. We then explore and test the limitations of our methodology for a model system. We use our method to predict distances between dimer pairs in both two and three dimensions.  Additionally, for each pair we compute a Lennard-Jones energy, and demonstrate that our methodology can predict this energy to a high degree of accuracy. We perform interpolation and extrapolation tests, and vary different controllable parameters to further understand our methodology. Using this success as motivation, we then use our new methodology to predict the total energy according to DFT within the generalized gradient approximation to a very high accuracy for various hexagonal lattices: graphene, boron nitride, and graphene-boron nitride heterostructures.  Our method is also able to predict energies for structures containing vacancies and Stone-Wales (five-seven) defects. 

\section{Methods}
\subsection{Input representation}
Since our method uses a deep CNN which exploits spatial structure in the input data structure, we decided to represent our atomic configurations as the approximate nuclear potential evaluated on a real-space mesh.  While a Coulomb potential is the initial obvious choice, we used an atom-centered Gaussian representation to avoid the diverging Coulomb singularity.  We evaluate our function on a real space grid with the value at point $(x, y, z)$ given by:
\begin{equation}
  \label{image_eqn}
  \begin{array}{cc}
    V(x,y,z) = \sum_{i=1}^{N} Z_i \exp\left(-\frac{[(x - x_i)^2 + (y - y_i)^2 + (z - z_i)^2]}{2\gamma^2}\right)
    \end{array}
\end{equation}

where $x_i, y_i, z_i$ are the coordinates of atom $i$ with atomic number $Z_i$.  We chose $\gamma = 0.2$ \AA~ as the width of the Gaussian peaks, consistent with Brockherde \emph{et al.} \cite{Brockherde}.  For the two dimensional images, the $z$ coordinate is not included. We note that the choice of $V$ is arbitrary, so long as it is consistent, and the relative peaks of the atoms are maintained.
\subsection{The datasets}
The datasets we generated consisted of two groups: dimer pairs and hexagonal sheets.
For the dimer dataset, we randomly generated two position vectors, $\vec {r}_1$ and $\vec {r}_2$, so that the distance between the two points $r_{12} = |\vec {r}_1 - \vec {r}_2|$ was within a specified range of values (e.g. $1.0\leq r_{12}\leq 2.0$ \AA). To accomplish this, we place the first atom down randomly, and then place down the second atom so that the distance between the two is maintained. The angle between the two position vectors is also randomly chosen when placing the second atom. To make sure that we do not reproduce a previously generated image, we declared a minimum dimer molecule overlap distance of 0.01 \AA. With this overlap distance specified, we  then initialized arrays for every grid point on a discretized grid ($0\leq x \leq 10$ \AA) with a $\Delta x=0.01$ \AA~ spacing. When we generated positions for the first dimer molecule, we found the arrays associated with the dimer coordinates, and added the index of this molecule to these arrays. When we generated additional positions for dimer molecules we first found the arrays associated with these proposed coordinates and checked to see if there is an intersection of indices between these two arrays as well as the arrays associated with neighbouring grid points. If there was an intersection, we proposed new coordinates. If not, we recorded the index in the arrays and continued generating new positions.  Using these coordinates, we evaluate Eq. (\ref{image_eqn}) on a $256\times 256$ grid-point mesh for 2D, and a $64\times 64\times 64$ grid-point mesh for 3D. In both cases, the real-space length of one side of the mesh is \mbox{$10$ \AA}.  Both dimer atoms have the same atomic number: $Z_1 = Z_2 = 1$.  We generated 500,000 2D images and 100,000 3D images. For each dimer image, we recorded two labels on which we would ultimately train the deep neural network: the distance $r_{12}$ and the Lennard-Jones energy 
\begin{equation}
  \label{lj}
U(r_{12})=4\epsilon\left[\left(\frac{\sigma}{r_{12}}\right)^{12} - \left(\frac{\sigma}{r_{12}} \right)^6\right],
\end{equation}
where we take $\sigma=1$ \AA~ and $\epsilon = 1$. Note that one can construct a CNN to predict both the distance and energy simultaneously.

To generate the configurations of hexagonal sheets, we performed Born-Oppenheimer molecular dynamics (BOMD) using DFT on systems of graphene, hBN, and a graphene-hBN heterostructure, all consisting of 60 atoms. 
Additionally, we generated datasets with single point defects, as well as Stone-Wales defects. The calculations were carried out using VASP \cite{VASP1,VASP2,VASP3,VASP4}, with the PBE exchange correlation functional \cite{perdew1996generalized}. All of the supercell dimensions were $12.53\times13.02\times10.0$ \AA, and the atoms were constrained along the $z$-axis at $z=0.0$ to allow for a two-dimensional treatment. We used a Nos\'e-Hoover thermostat of 1000 K, a plane wave kinetic-energy cutoff of 500 eV, and a ${\bf k}$-point grid of $2\times2\times1$ centred about the $\Gamma$ point. For the MD, a timestep of 11.3 a.u. was used in the simulations. For each type of hexagonal sheet, the training set was generated by running many independant sets of MD for 0.15 picoseconds (approximately 550 steps). After the completion of one MD run, the final coordinates were randomly translated in the $x$ and $y$-directions, and the velocities were reinitialized using a Maxwell-Boltzmann distribution. This process was repeated until approximately 9 picoseconds per hexagonal structure was generated. In total, we generated 269,016 images in our generation process. Using the coordinates of the atoms from the molecular dynamics frames, the training images were generated using Eq. (\ref{image_eqn}), summing over all atoms in the unit cell.  The atomic numbers, $Z_i$ were used so that atoms of higher atomic number had a larger Gaussian peak. The pixels were wrapped to obey periodic boundary conditions.

\begin{figure}[h!]
  \centering
  \includegraphics[width=\linewidth]{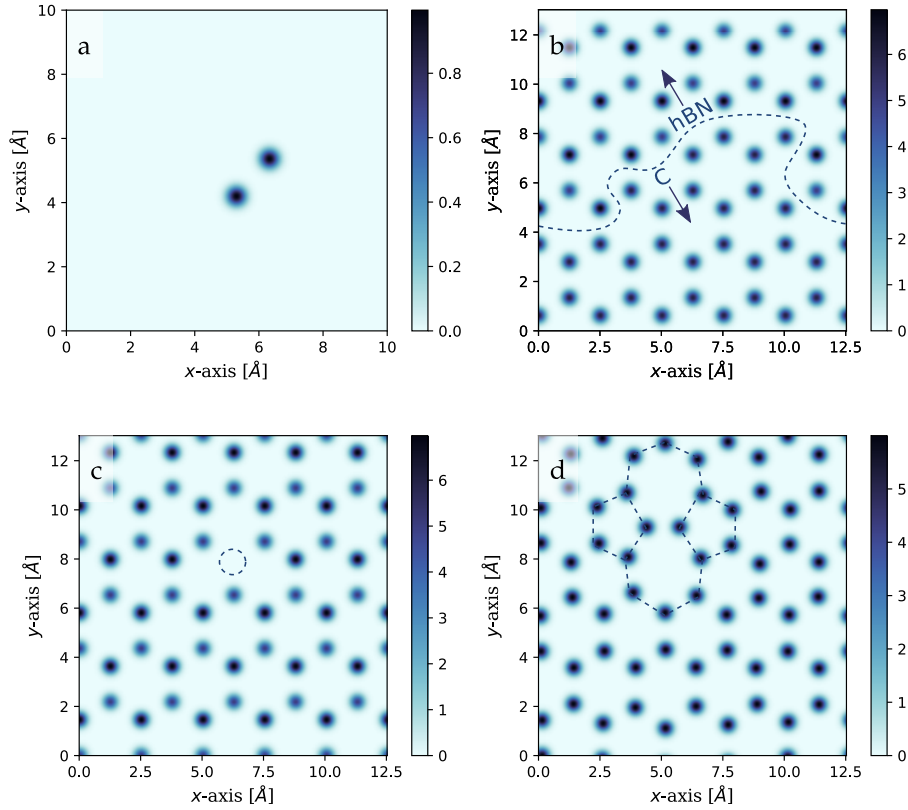}
  \caption{\label{sample_images} Images used as input to the CNNs. (a): A dimer molecule image. (b): An image of a graphene-hBN heterostructure. The dashed lines divides the hBN structure and graphene. (c): An image of hBN with a N atom removed (single point defect, indicated with dashed line). (d): An image of graphene with a Stone-Wales defect in the region indicated by the dashed lines.}
\end{figure}

After the generation of all datasets, they were then split randomly so that 70\% comprised a training set and 30\% comprised a testing set. While training, 10\% of the training set was used as a validation set. All of the testing sets were only used to compute errors, and were not be accessible to the neural network during the training process.  Some example images of the input datasets are shown in Figure \ref{sample_images}. 

\begin{figure}[h!]
  \centering  
\includegraphics[width=0.5\linewidth]{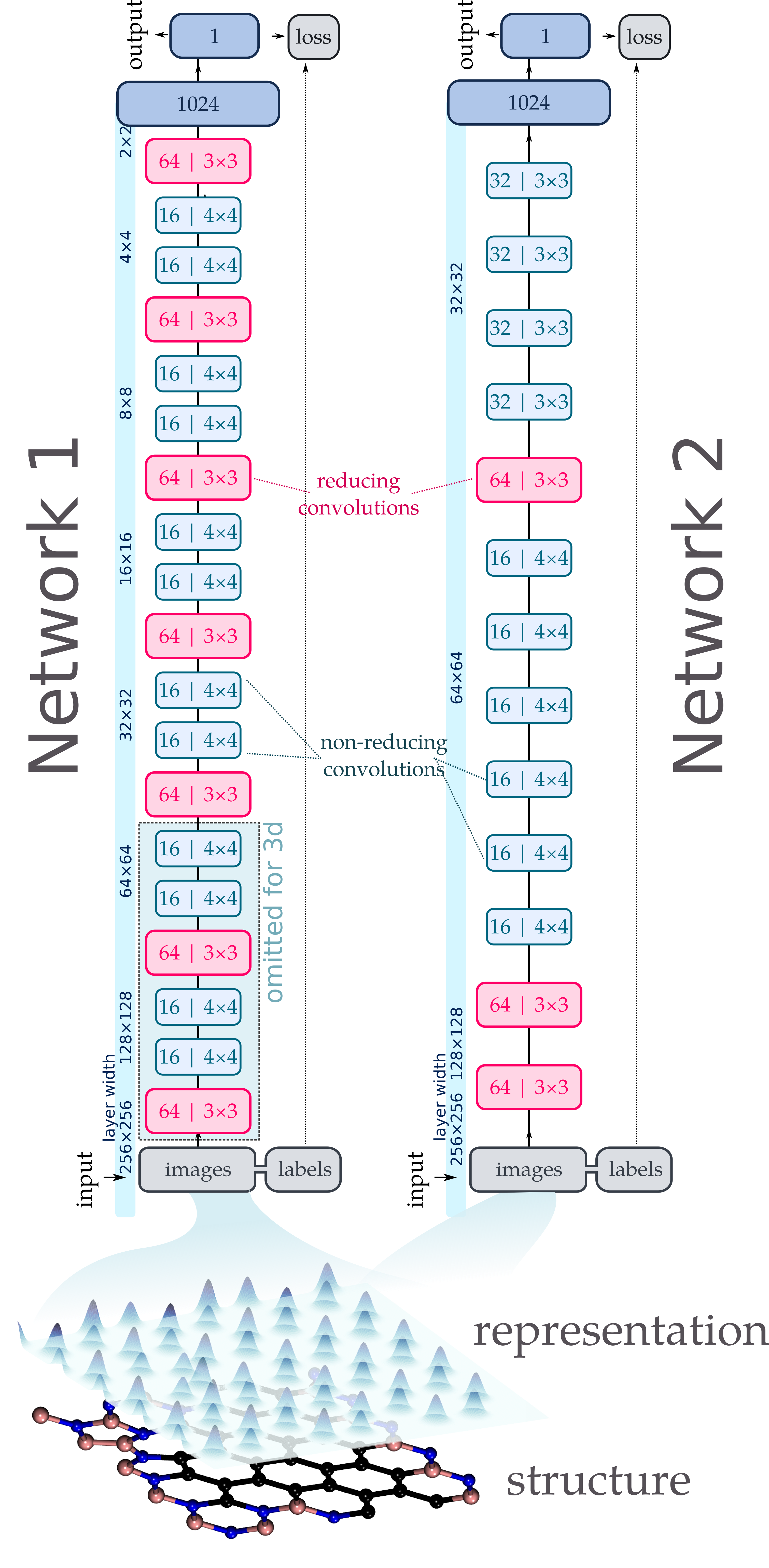}

\caption{\label{networks} Schematic description of the methodology. The atomic structures are first mapped to a Gaussian surface described by Equation \ref{image_eqn}. These images are then passed into the CNNs. Network 1 was used for all dimer models, and Network 2 was used for the hexagonal structures. Reducing convolutional layers operate with a stride of 2 in each direction, and reducing convolutional layers operate with unit stride.  In the diagram, the number of filters and the filter dimension (for two dimensions) are shown. These networks were optimized in parallel on graphical processing units (GPUs) using Tensorflow \cite{abadi2016tensorflow}.}
\end{figure}

\subsection{The CNNs}

We used two relatively deep neural network architectures, shown in Figure \ref{networks}, comprised of a combination of reducing (stride 2) and non-reducing (stride 1) convolutional layers (CLs). CLs consist of an array of kernels that operate on images. The kernel sizes of CLs determine the number of parameters that are optimized during the training process. When using a reducing CL, the images that are output from the CL will have reduced dimensionality. When using a non-reducing CL, the images that are output from the CL will have the same dimensionality. We do not use dropout, or any sort of pooling in our network architectures. Network 1 has the identical architecture used in \cite{mills2017deep}, and was used for learning the dimer distances and energies.  For the 3D models, three dimensional filters were used (e.g. $4\times 4\times 4$ instead of $4\times 4$), and certain layers (shown in Figure \ref{networks}) were omitted to accommodate the smaller input dimension. Within the convolutional layers, we used zero padding when applying convolutional kernels to the edge pixels. This allows for the dimensionality of the convolved images to remain the same for non-reducing layers, and to decrease exactly by half in the reducing layers. Network 2 was modified for better performance on the hexagonal sheets. When constructing this model, our guiding principle was simplicity. We used trial and error in removing layers from Network 1 until we were able to reach a certain accuracy in our predictions. It is very likely that a more optimized architecture (found through a service like SigOpt \cite{dewancker2016evaluation}) could result in even better performance.  In order to include periodic boundary conditions in our method, we used four shifted and wrapped copies of each image during training. The four copies constitute a shift such that each of the four original boundaries appear at the centre of the image in at least one of the copies. This leads to a network topology with 4 neural networks (with the same weights) being training concurrently, a fully connected layer with 1024 neurons to combine the output layers of the 4 networks, and final fully connected layer that outputs the prediction. We used rectified linear units (ReLU) for all activations, and trained using the Adam optimization scheme \cite{kingma2014adam} to minimize the mean-squared error between the correct energy/distance and the CNN output. The use of ReLU activations mean that the computed gradients with respect to weights will be constant, which improves the efficiency of the backpropagation algorithm and is less demanding to evaluate than the sigmoid function (since the derivatives of the activation functions with respect to weights are constant). For the dimer models, we trained for 500 epochs with a constant learning rate of $10^{-4}$. For the model making energy predictions of the hexagonal sheets we trained for 300 epochs with a learning rate of $10^{-5}$, and then dropped the learning rate to ${10}^{-6}$ before training for another 200 epochs. All of the models were trained using TensorFlow \cite{abadi2016tensorflow}.

\section{Results}

\begin{figure}[h!]
 \centering
 \includegraphics[width=\linewidth]{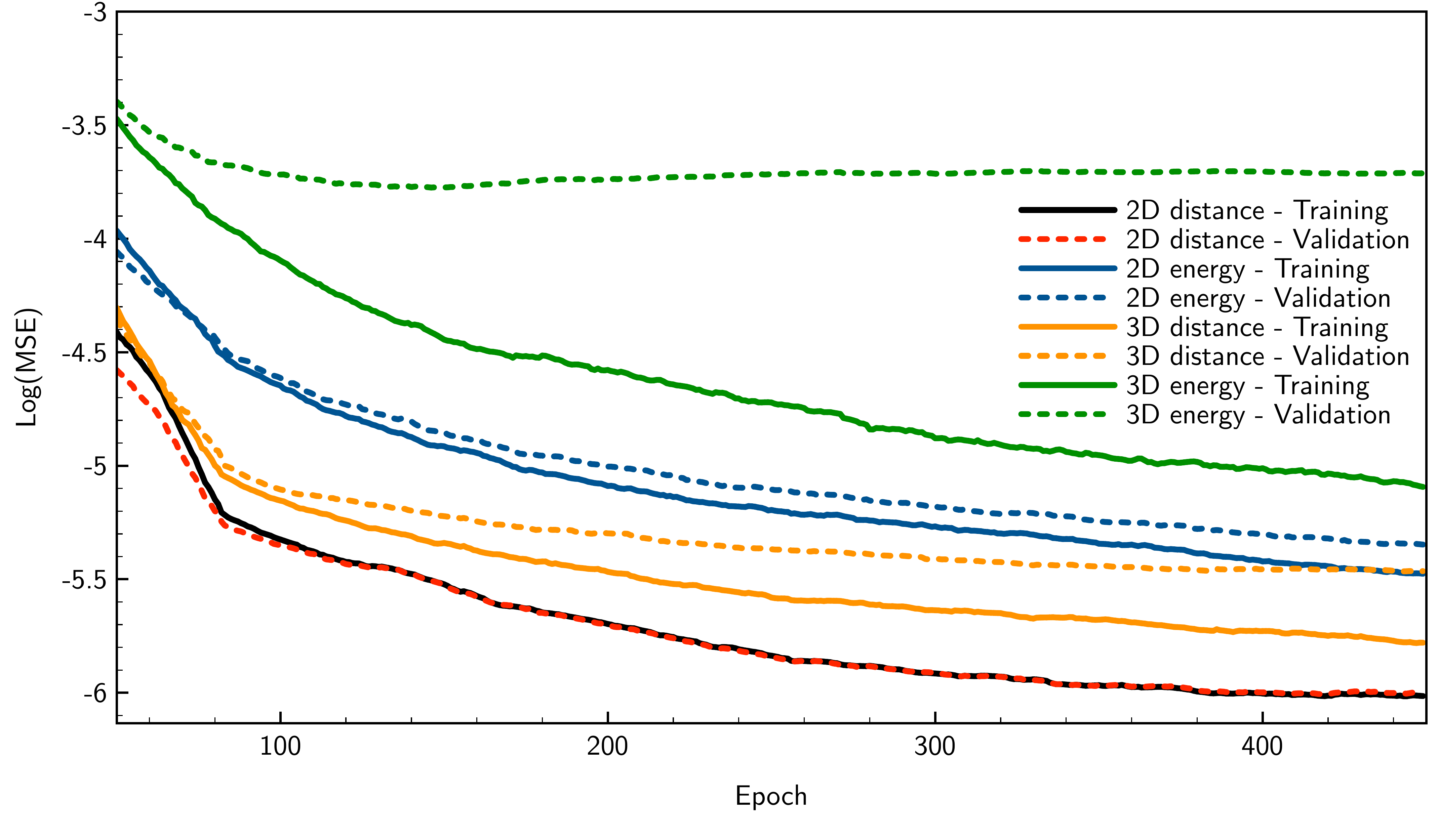}
 \caption{\label{convergence} Running averages (with a window of 50) of the mean squared errors (loss) recorded during training.  The models are the following: A 2D model for predicting distances in the range $1\leq r_{12}\leq 3$ (2D distance), a 3D model for predicting distances in the range $1 \leq r \leq 3$ (3D distance), a 2D model for predicting energies in the range $-1\leq U(r_{12}) \leq 0$ (2D energy), and a 3D model for predicting energies in the range $-1\leq U(r_{12})\leq 0$ (3D energy). The units of the loss function is in \AA${}^2$ for the distance models, and $\epsilon^2$ for the energy models.}
\end{figure}

\subsection{Dimer pairs}
\begin{table}
\begin{center}
  \scriptsize
\begin{tabular}{c|c|c|c}
  Grid size & System description & MAE (testing set) & MSE (validation set)\\\hline
  $256\times256$ & $1\leq r_{12} \leq 3$ & $1.58\times10^{-3}$ \AA&  $5.52\times10^{-6}$ \AA${}^2$ \\
  $256\times256$ & $1\leq r_{12} \leq 2$ & $9.99\times10^{-4}$ \AA &  $1.48\times10^{-6}$ \AA${}^2$ \\
  $256\times256$ & $2\leq r_{12} \leq 3$ & $9.72\times10^{-4}$ \AA &  $1.47\times10^{-6}$ \AA${}^2$ \\
  $256\times256$ & $1\leq r_{12} \leq 1.67$ \& $2.33\leq r_{12} \leq 3$ & $3.52\times10^{-3}$ \AA &  $3.48\times10^{-6}$ \AA${}^2$ \\

  $64\times64\times64$ & $1\leq r_{12} \leq 3$ & $3.21\times10^{-3}$ \AA & $1.64\times10^{-5}$ \AA${}^2$ \\
  $64\times64\times64$ & $1\leq r_{12} \leq 2$ & $1.58\times10^{-3}$ \AA &  $4.20\times10^{-6}$ \AA${}^2$\\
  $64\times64\times64$ & $2\leq r_{12} \leq 3$ & $1.72\times10^{-3}$ \AA & $4.77\times10^{-6}$ \AA${}^2$ \\
  $64\times64\times64$ & $1\leq r_{12} \leq 1.67$ \& $2.33\leq r_{12} \leq 3$ & $2.93\times10^{-3}$ \AA & $1.50\times10^{-5}$ \AA${}^2$ \\

  $256\times256$ & $-1\leq U(r_{12}) \leq 0$ & $1.41\times10^{-3}$ $\epsilon$ & $5.82\times10^{-6}$ $\epsilon^2$  \\
  $256\times256$ & $-0.5\leq U(r_{12}) \leq 0$ & $1.62\times10^{-3}$ $\epsilon$ & $5.29\times10^{-6}$ $\epsilon^2$ \\
  $256\times256$ & $-1\leq U(r_{12}) \leq -0.5$ & $1.62\times10^{-3}$ $\epsilon$ & $2.33\times10^{-6}$ $\epsilon^2$  \\
  $256\times256$ & $-1\leq U(r_{12}) \leq 0.67$ \& $-0.33\leq U(r_{12}) \leq 0$  & $1.41\times10^{-3}$ $\epsilon$ & $3.72\times10^{-6}$ $\epsilon^2$  \\
  
  $64\times64\times64$ & $-1\leq U(r_{12}) \leq 0$ & $5.11\times10^{-3}$ $\epsilon$ & $1.91\times10^{-4}$ $\epsilon^2$  \\
  $64\times64\times64$ & $-0.5\leq U(r_{12}) \leq 0$ & $1.14\times10^{-2}$ $\epsilon$ & $3.19\times10^{-4}$ $\epsilon^2$  \\
  $64\times64\times64$ & $-1\leq U(r_{12}) \leq 0.67$ \& $-0.33\leq U(r_{12}) \leq 0$  & $9.43\times10^{-3}$ $\epsilon$ & $2.05\times10^{-4}$ $\epsilon^2$  \\
  
  $32\times32$ & $-1\leq U(r_{12}) \leq 0$ & $4.99\times10^{-1}$ $\epsilon$ & $3.31\times10^{-1}$ $\epsilon^2$  \\
  $64\times64$ & $-1\leq U(r_{12}) \leq 0$ & $1.70\times10^{-3}$ $\epsilon$ & $4.02\times10^{-6}$ $\epsilon^2$  \\
  $128\times128$ & $-1\leq U(r_{12}) \leq 0$ & $1.06\times10^{-3}$ $\epsilon$ & $1.67\times10^{-6}$ $\epsilon^2$  \\
  
    
  $256\times256$ & $1\leq r_{12} \leq 2$ (Random forest) & $1.68\times10^{-2}$ \AA &  - \\
  $256\times256$ & $1\leq r_{12} \leq 2$ (Kernel ridge regression - linear) &  $1.24\times10^{-1}$ \AA & -  \\
    $256\times256$ & $1\leq r_{12} \leq 2$ (Kernel ridge regression - Gaussian) &  $2.47\times10^{-1}$ \AA & -  \\
  $256\times256$ & $1\leq r_{12} \leq 2$ (Multilayer perceptron) & $1.24\times10^{-1}$ \AA & -  \\
  
  $256\times256$ & Hexagonal sheets & 0.0119 eV & $3.11\times10^{-4}$ eV${}^2$
  
\end{tabular}
\end{center}
\caption{\label{table} Mean absolute errors and mean squared errors of various systems with their corresponding test and validation sets. The mean squared errors are taken from epoch 500.}
\end{table}

We demonstrate the ability of the CNNAS approach at predicting the separation distance between dimer pairs and the Lennard-Jones energies using the following four models:
\begin{enumerate}
\item a 2D model for predicting distances in the range $1\leq r_{12}\leq 3$,
\item a 3D model for predicting distances in the range $1 \leq r_{12} \leq 3$,
\item a 2D model for predicting energies in the range $-1\leq U(r_{12}) \leq 0$, and
\item a 3D model for predicting energies in the range $-1\leq U(r_{12})\leq 0$.
\end{enumerate}

In Figure \ref{convergence} we plot the running average of the training and validation loss as a function of epoch (one time through the training set) for each model.  After five hundred epochs, we see that the CNNs are converged (Figure \ref{convergence}). When comparing the models in Table \ref{table}, we find that the 2D models outperform the 3D models in all cases. This can be attributed to the amount of training data provided to both models. Since the 2D models are significantly faster to train, we were able to provide the deep neural network with more training examples. To investigate this, we trained a 3D model with 500,000 images and found that the MAE decreases by 85\%.  Additionally, the difference in training data resolution plays a role. The pixel density in the 2D dimer dataset is higher than in the 3D dimer dataset.  When testing a 2D energy predicting model on a $64\times64$ grid ($2.44\times10^{-2}$ \AA${}^2$ pixel area) rather than a $256\times256$ grid ($1.53\times10^{-3}$ \AA${}^2$ pixel area), the MAE dropped to $1.70\times10^{-3}~\epsilon$ (21\% difference). To further investigate the grid sizes, we calculated the MAEs of test sets for 2D energy models with identical dataset sizes but differing grid sizes. We changed the grid sizes from $32\times32$ to $128\times128$ in multiples of 2, and found that an energy predicting model performed best with a $128\times128$ grid. In addition to investigating the grid size, we also carried out experiments where we upscaled lower dimensional images (e.g.  $64\times64$ grids) into $256\times256$ grids. This was done by replicating the pixels in the lower dimensional image. When upscaling from $64\times64$ to $256\times256$, each pixel would be replicated 16 times (4 in the horizontal direction and 4 in the vertical). We found a similar MAE when comparing the upsampled images and the original images. A non-upscaled $256\times256$ grid contains enough information for the DNN to make accurate predictions. A $256\times256$ resolution image is sufficient to generate an accurate DNN for our atomistic systems. 

\begin{figure}[h!]
 \centering
 \includegraphics[width=\linewidth]{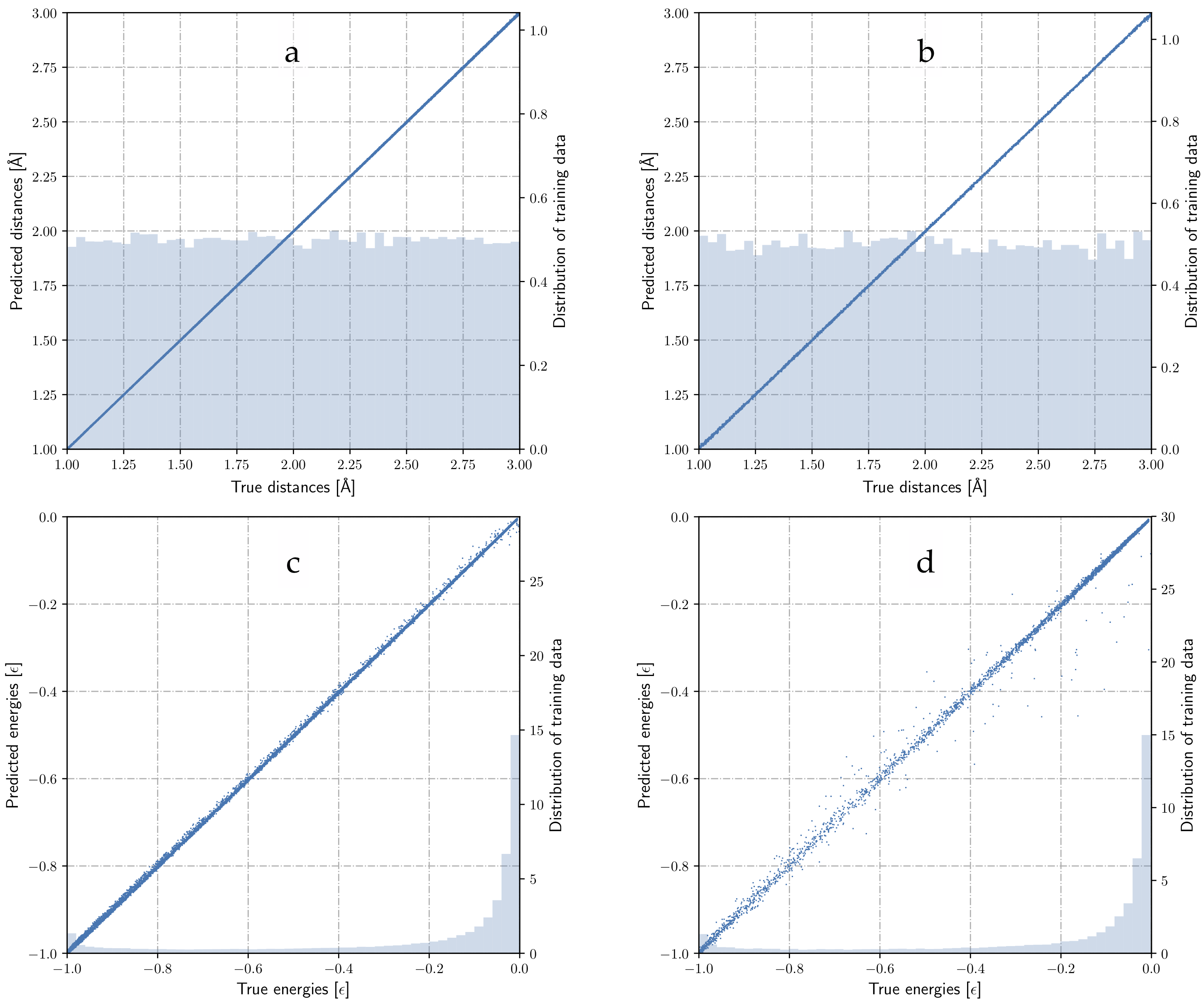}
 \caption{\label{main_pred_vs_true} Predicted versus true plots for various models. (a) The 2D model for predicting distances in the range $1\leq r_{12}\leq 2$. (b) The 3D model for predicting distances in the range $1 \leq r \leq 2$. (c) The 2D model for predicting energies in the range $-1\leq U(r_{12}) \leq 0$. (d) the 3D model for predicting energies in the range $-1\leq U(r_{12})\leq 0.0$. The light blue distribution in the background of all plots shows the distribution of data used in the training of the models.}
\end{figure}

The differences between the different models can also be seen in Figure \ref{main_pred_vs_true}, where we plot the predicted versus true values for these models as well as the distribution of input data. For the 3D predicted versus true scatter plots, there is much more variance in the distribution of points in comparison to the 2D models which is due to less training data and the pixel density, as discussed above. When comparing distance models with the energy models, we can visually see that the variance in the energy distributions are higher than the distance models. To investigate this systematically, we trained 3 independent models:

\begin{enumerate}
\item A 2D model trained on a harmonic function $U(r) = (r-2.0)^2$ for $1.0\leq r \leq 3.0$.
\item A 2D model trained on the function $U(r) = \frac{1}{r^{12}}$ for $1.0\leq r \leq 3.0$.
\item A 2D energy model with uniform sampling.
\end{enumerate}

For model trained on the harmonic function and the function $r^{-12}$, we found that the variance in the predicted versus true plots was comparable to the variance in the distance models shown in Figure \ref{main_pred_vs_true}. The CNNs are capable of handling both the non-linearity and the rapid change in the Lennard-Jones function as $r\rightarrow0$. For the model trained on uniformly sampled energies, we found a 20\% decrease in the root mean squared error (RMSE) when comparing the uniformly sampled model with the non-uniformly sampled model (i.e the model shown in Figure \ref{main_pred_vs_true}). We therefore conclude that the distance models perform better simply due to our sampling technique in the data generation process. For all of the dimer models in this manuscript, we uniformly sampled the distance, not the energy. When examining the distribution of input distances and energies, one can clearly see the non-uniformity in the distribution of energies for both the 2D and 3D models. 


To compare with classic machine learning methods, we also performed tests using a multilayer perceptron (MLP), kernel ridge regression (KRR), and random forests (RF) for 50,000 distances in the range $1\leq r_{12}\leq2$. To perform these tests, we used the scikit-learn \cite{pedregosa2011scikit} framework in Python. For KRR, we tried a linear and Gaussian kernel. For the  linear kernel we used an alpha value of 1, and a degree 3 polynomial with a coefficient of 1. For the Gaussian kernel, we used $\gamma=1,2,4$ and $16$. We found that all of these parameters gave similar results for the MAE on the test set, which is reported in Table \ref{table}. For the MLP, we used 2 hidden layers consisting of 10 neurons, ReLU activation functions, the Adam optimization scheme, a learning rate of 0.001, and 200 epochs. For the RF model, we used 200 estimators (trees), the mean squared error to measure the quality of a split, and the maximum number of features was $256\times256$. The input to these models was the raw flattened images. RF performed best, with a MAE of 0.0168 \AA, while MLP and KRR performed similarly, with a MAE of 0.124 \AA. We found that the DNN outperforms all of these models with a MAE of $9.99\times10^{-4}$ \AA. The CNNAS approach for dimer molecules with raw data in both two and three dimensions is extremely accurate. It should be noted that the traditional machine learning models can be improved with feature engineering and parameter optimization. As an example, Brockherde \emph{et al.} \cite{BVLT17} used KRR and were able to make energy predictions for a H${}_2$ molecule within chemical accuracy with only 200 training examples. A DNN can only perform well with many training examples due to the large number of tuneable parameters. In the low volume data domain (e.g. only a few hundred training examples) a traditional machine learning model would be a more suitable choice. In the high volume data domain, DNNs are the suitable choice. The features are learned from the raw data, which avoids the feature engineering stage of constructing a machine learning model using a traditional approach.

\begin{figure}[h!]
 \centering
 \includegraphics[width=\linewidth]{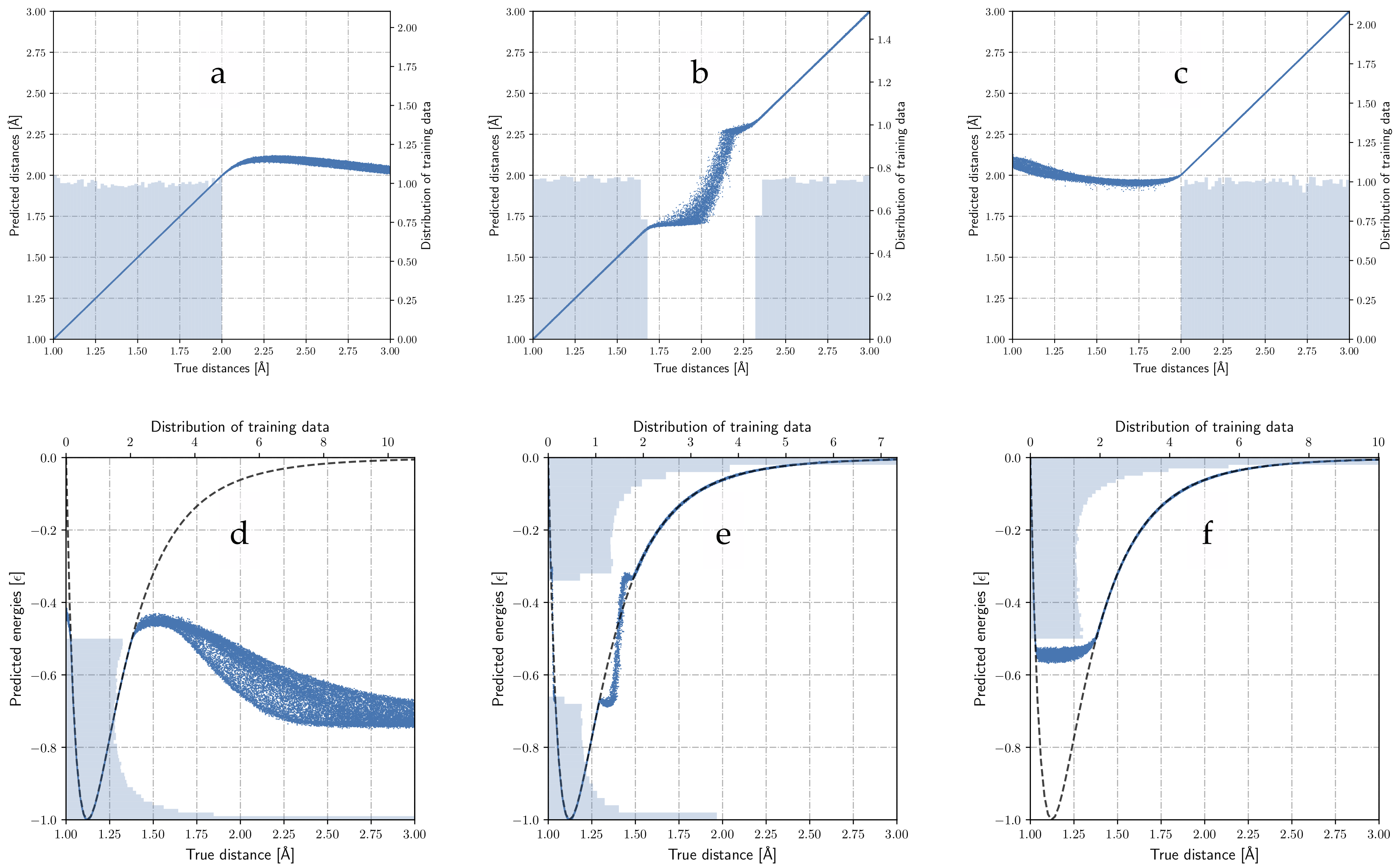}
 \caption{\label{interp_vs_extrap} Predicted distance versus true distance (labelled a, b, and c) and predicted energy versus true distance (labelled d, e, and f) plots for the interpolation and extrapolation experiments. (a), (b), and (c) are the true distance versus predicted distance plots for the 2D distance models. (d), (e), and (f) are predicted energy versus true distance for to the 2D energy models. (a), (c), (d), and (f) correspond with extrapolation, and (b) and (e) correspond to interpolation, all described in subsection \ref{interp_and_extrap}. The light blue distribution shown in the background of all plots gives the distribution of training data (a, b, and c correspond to distance distributions and d, e, and f correspond to energy distributions). The dotted grey lines in (d), (e), and (f) is the model Lennard-Jones curve, described by Equation \ref{lj}.}
\end{figure}

\subsection{Interpolation and Extrapolation}
\label{interp_and_extrap}
To investigate how well the CNNAS handles interpolating and extrapolating distances and energies, we constructed 2D and 3D models where we excluded training examples within a certain range. For distances we define the total range to be $1\leq r_{12}\leq3$, and for energies we define the total range to be $-1\leq U(r_{12})\leq0$. When we perform an extrapolation test, we task the model to predict values that are greater or less than the range of the original dataset the model was trained on. When we perform an interpolation test, we task the model to predict values inside of the range it has been trained on, but in regions where it has not seen training examples. Although outside of the range it has been trained on, this range is within the minimum and maximum values of the original training dataset. We then trained 2D and 3D models for three different distance and energy ranges within their respective total ranges. Looking to Figure \ref{interp_vs_extrap}, we clearly see that all of the models fail when extrapolating. The models are only capable of predicting values within the range they have been trained on. Therefore, when the models are predicting values outside of this range, they return values on the endpoints of the ranges. To avoid this extrapolation issue, one must be aware that the model should see as large of a range as possible to make predictions for a general system. We found that in the interpolation tests, the model's ability to interpolate was also poor. Although these models were not trained on certain regions within the total ranges, they were able to make predictions within these regions. When tasked with interpolation, the MAE of the test set for the 2D distance interpolation model was $3.43\times10^{-2}$ \AA, which is a 1070\% increase in comparison to MAE of the original corresponding test set. The original corresponding test set does not task the network to predict values outside of the range it has originally been trained on. The original test set has the same range as the training set. All of the error comes from the interpolation region. Similar effects were observed for the 2D and 3D distance models.


\subsection{Hexagonal sheets}
\begin{figure}[h!]
 \centering
 \includegraphics[width=\linewidth]{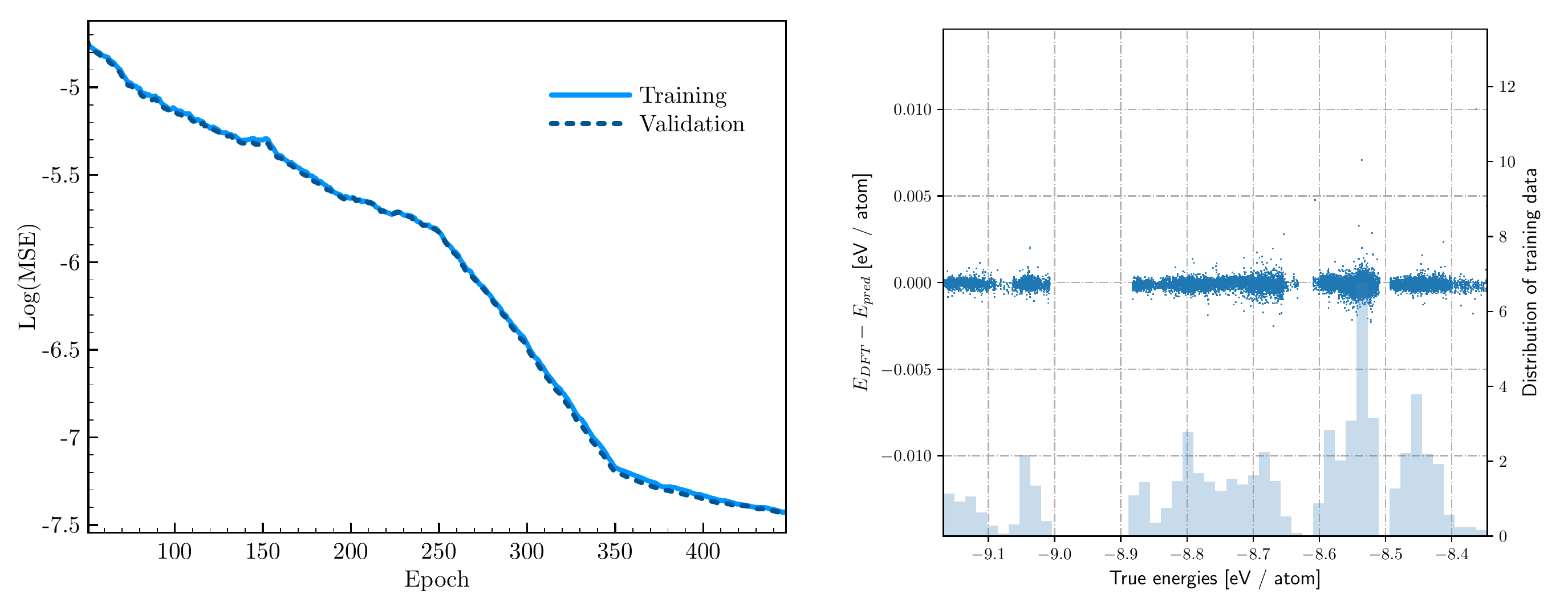}
 \caption{\label{pred_vs_true_dft} Left: Running average (with a window size of 50) of the training and validation loss during the training process. Right: DFT energy per atom minus the predicted energy per atom for the DFT model. The light blue distribution in the background gives the distribution of data used to train the model. The gap in the distribution arises from the absence of DFT energies within that particular range. At 300 epochs, the learning rate was lowered from $10^{-5}$ to $10^{-6}$.}
\end{figure}
After concluding that our methodology allows for a successful DNN model to predict dimer distances and energies, we then moved on to much more complex many-body systems. As mentioned previously, we created an input data set for a DNN model by generating first principles molecular dynamics for a graphene sheet, a hBN sheet, and a graphene-hBN heterostructure. Within these structures, we also created either single point defects by removing one atom, or Stone-Wales defects by deforming the crystal lattice. After collecting the molecular dynamics data, we converted the molecular dynamics trajectories to images using Equation \ref{image_eqn}, and combined all of the images together into one data set. We first trained using network architecture 1 from Figure \ref{networks}, but we found that the loss as a function of epoch did not decrease exponentially. Due to the increase of information (or number of atoms) within the images, the number of reducing convolutional layers eliminated information in the network necessary for making accurate energy predictions. This led to our choice of network architecture 2, also seen in Figure \ref{networks}. This network architecture has fewer reducing convolutional layers, which allows for more information to be transmitted to the final fully connected layer. We found that this aided in the prediction process of DFT energies. We found the loss to decrease exponentially, indicating the successful training of this model. In Figure \ref{pred_vs_true_dft}, we can clearly see that the network does exceptionally well at predicting DFT energies for trajectories it had not seen before. The MAE of the test set was an impressive 0.0119 eV for total energies, or 0.198 meV / atom. Not only is the accuracy exceptional, but the computational cost was minimal. We were able to calculate approximately one hundred thousand total energies on the order of minutes.\\

\begin{figure}[h!]
 \centering
 \includegraphics[width=\linewidth]{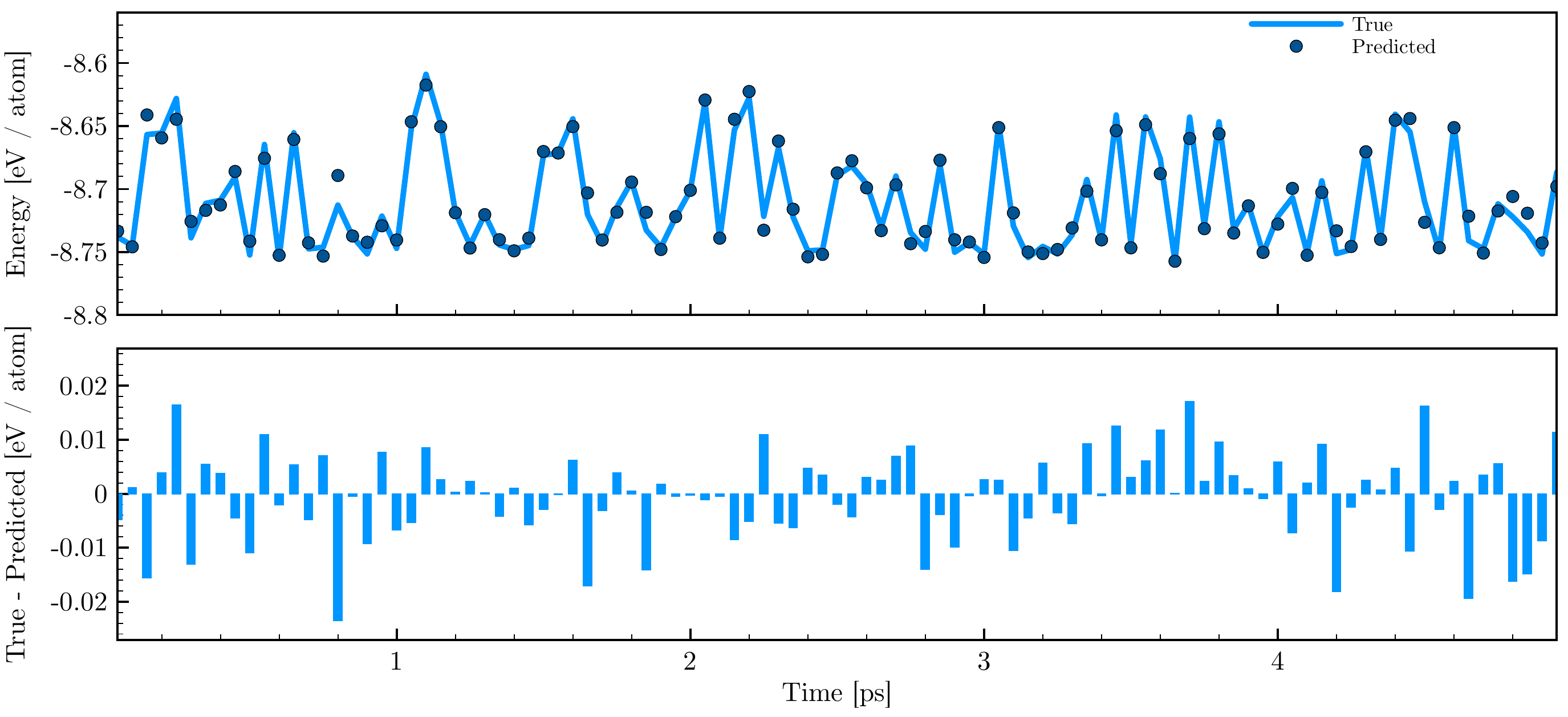}
 \caption{\label{md_preds} Top: True DFT and predicted energies per atom as a function of time for the hBN MD for every 100 steps. Bottom: True DFT energy per atom minus the predicted energy per atom as a function of time for the hBN MD for every 100 steps.}
\end{figure}

The viability and extensibility of a machine learned model can be shown by evaluating the model on its corresponding testing set, but a more rigorous test for a model is using it in practice. To further demonstrate the extensibility of our model, we performed additional MD calculations of the hexagonal-boron nitride surface and calculated predicted energies using the model for every 100 steps of the MD. For this newly generated MD we used the same parameters as before when generating the data, but we used a time step of 20.7 atomic units. We collected 5 ps of data for the new predictions. Looking to Figure \ref{md_preds}, we plot the DFT and predicted energies as a function of time for every 100 MD steps. The MAE of the new predictions is 6.85 meV / atom. The maximum absolute error predicted for the MD trajectory is 38.8 meV / atom. This additional test further confirms the viability and extensibility of our machine learned model.

\section{Conclusion}
We have shown that our new methodology can be used successfully for predicting atomic distances and energies for dimer molecules as well as DFT energies for 2D hexagonal sheets. For dimer molecules, we found that our method is most accurate in 2D, which can be attributed to the increase of training data, and increased grid spacing in comparison to the 3D models. When testing the limits of our method, we found that the models are limited by the range of data they have been trained on. When extrapolating, the models predict values on the boundaries of the ranges they have been trained on. When testing the models' ability to interpolate, the models also make poor predictions. Although the models have not seen any data within the interpolation regions, they are still able to make predictions within the space spanned by the minimum and maximum values of the training set. Lastly, and most importantly, we found that CNNAS perform exceptionally well when tasked to predict DFT energies for a variety of hexagonal surfaces. The MAE we found was 0.0119 eV for the total energies, or 0.198 meV / atom for the test dataset. In addition, when the model was tasked to calculate energies of a new MD trajectory for hBN, it also succeeded with a MAE of 6.85 meV / atom. 
\section{Acknowledgement}
The authors acknowledge NSERC, OGS, and SOSCIP for funding and computational resources, and NVIDIA for a faculty hardware grant.

%
%
%




\bibliographystyle{elsarticle-num}


%
%
%

\bibliography{km_no_url,refs_no_url}

\begin{thebibliography}{10}
\expandafter\ifx\csname url\endcsname\relax
  \def\url#1{\texttt{#1}}\fi
\expandafter\ifx\csname urlprefix\endcsname\relax\def\urlprefix{URL }\fi
\expandafter\ifx\csname href\endcsname\relax
  \def\href#1#2{#2} \def\path#1{#1}\fi

\bibitem{Born1927}
M.~Born, R.~Oppenheimer, {Zur Quantentheorie der Molekeln}, Annalen der Physik
  389~(20) (1927) 457--484.
\newblock \href {http://arxiv.org/abs/1206.4239} {\path{arXiv:1206.4239}},
  \href {http://dx.doi.org/10.1002/andp.19273892002}
  {\path{doi:10.1002/andp.19273892002}}.

\bibitem{Equations1965}
W.~Kohn, L.~J. Sham, {Self-consistent equations including exchange and
  correlation effects}, Physical Review 140~(4A).
\newblock \href {http://arxiv.org/abs/PhysRev.140.A1133}
  {\path{arXiv:PhysRev.140.A1133}}, \href
  {http://dx.doi.org/10.1103/PhysRev.140.A1133}
  {\path{doi:10.1103/PhysRev.140.A1133}}.

\bibitem{Ratcliff2016b}
L.~E. Ratcliff, S.~Mohr, G.~Huhs, T.~Deutsch, M.~Masella, L.~Genovese,
  {Challenges in Large Scale Quantum Mechanical Calculations}, WIREs
  Computational Molecular Science 7~(1) (2016) e1290.
\newblock \href {http://arxiv.org/abs/1609.00252} {\path{arXiv:1609.00252}},
  \href {http://dx.doi.org/10.1002/wcms.1290} {\path{doi:10.1002/wcms.1290}}.

\bibitem{Mercado2016}
R.~Mercado, B.~Vlaisavljevich, L.-C. Lin, K.~Lee, Y.~Lee, J.~A. Mason, D.~J.
  Xiao, M.~I. Gonzalez, M.~T. Kapelewski, J.~B. Neaton, B.~Smit, {Force Field
  Development from Periodic Density Functional Theory Calculations for Gas
  Separation Applications Using Metal–Organic Frameworks}, The Journal of
  Physical Chemistry C 120~(23) (2016) 12590--12604.
\newblock \href {http://dx.doi.org/10.1021/acs.jpcc.6b03393}
  {\path{doi:10.1021/acs.jpcc.6b03393}}.

\bibitem{Cherukara2016}
M.~J. Cherukara, B.~Narayanan, A.~Kinaci, K.~Sasikumar, S.~K. Gray, M.~K. Chan,
  S.~K. R.~S. Sankaranarayanan, {Ab Initio -Based Bond Order Potential to
  Investigate Low Thermal Conductivity of Stanene Nanostructures}, The Journal
  of Physical Chemistry Letters 7~(19) (2016) 3752--3759.
\newblock \href {http://dx.doi.org/10.1021/acs.jpclett.6b01562}
  {\path{doi:10.1021/acs.jpclett.6b01562}}.

\bibitem{Kulkarni2015}
A.~R. Kulkarni, D.~S. Sholl, {DFT-Derived Force Fields for Modeling Hydrocarbon
  Adsorption in MIL-47(V)}, Langmuir 31~(30) (2015) 8453--8468.
\newblock \href {http://dx.doi.org/10.1021/acs.langmuir.5b01193}
  {\path{doi:10.1021/acs.langmuir.5b01193}}.

\bibitem{Jaramillo-Botero2014}
A.~Jaramillo-Botero, S.~Naserifar, W.~A. Goddard, {General multiobjective force
  field optimization framework, with application to reactive force fields for
  silicon carbide}, Journal of Chemical Theory and Computation 10~(4) (2014)
  1426--1439.
\newblock \href {http://dx.doi.org/10.1021/ct5001044}
  {\path{doi:10.1021/ct5001044}}.

\bibitem{VanBeest1990}
B.~W.~H. {Van Beest}, G.~J. Kramer, R.~A. {Van Santen}, {Force fields for
  silicas and aluminophosphates based on ab initio calculations}, Physical
  Review Letters 64~(16) (1990) 1955--1958.
\newblock \href {http://dx.doi.org/10.1103/PhysRevLett.64.1955}
  {\path{doi:10.1103/PhysRevLett.64.1955}}.

\bibitem{Ponder2003}
J.~W. Ponder, D.~A. Case, {Force fields for protein simulations}, Advances in
  Protein Chemistry 66 (2003) 27--85.
\newblock \href {http://dx.doi.org/10.1016/S0065-3233(03)66002-X}
  {\path{doi:10.1016/S0065-3233(03)66002-X}}.

\bibitem{Hornak2006}
V.~Hornak, R.~Abel, A.~Okur, B.~Strockbine, A.~Roitberg, C.~Simmerling,
  {Comparison of multiple amber force fields and development of improved
  protein backbone parameters}, Proteins: Structure, Function and Genetics
  65~(3) (2006) 712--725.
\newblock \href {http://arxiv.org/abs/0605018} {\path{arXiv:0605018}}, \href
  {http://dx.doi.org/10.1002/prot.21123} {\path{doi:10.1002/prot.21123}}.

\bibitem{Waals1873}
van~der Waals, {De continuiteit van den gasen Vloeistoftoestand}, Ph.D. thesis
  (1873).

\bibitem{Perez2007}
A.~P{\'{e}}rez, I.~March{\'{a}}n, D.~Svozil, J.~Sponer, T.~E. Cheatham, C.~a.
  Laughton, M.~Orozco, {Refinement of the AMBER force field for nucleic acids:
  improving the description of alpha/gamma conformers.}, Biophysical journal
  92~(11) (2007) 3817--29.
\newblock \href {http://dx.doi.org/10.1529/biophysj.106.097782}
  {\path{doi:10.1529/biophysj.106.097782}}.

\bibitem{Allinger1989}
M.~J. Hackett, J.~A. McQuillan, F.~El-Assaad, J.~B. Aitken, A.~Levina, D.~D.
  Cohen, R.~Siegele, E.~A. Carter, G.~E. Grau, N.~H. Hunt, P.~A. Lay, {Chemical
  alterations to murine brain tissue induced by formalin fixation: implications
  for biospectroscopic imaging and mapping studies of disease pathogenesis},
  Journal of the American Chemical Society 111~(23) (1989) 8551--8566.
\newblock \href {http://dx.doi.org/10.1039/c0an00269k}
  {\path{doi:10.1039/c0an00269k}}.

\bibitem{Allinger1977}
N.~L. Allinger, {Conformational analysis. 130. MM2. A hydrocarbon force field
  utilizing V1 and V2 torsional terms}, Journal of the American Chemical
  Society 99~(25) (1977) 8127--8134.
\newblock \href {http://dx.doi.org/10.1021/ja00467a001}
  {\path{doi:10.1021/ja00467a001}}.

\bibitem{Weiner1984}
S.~J. Weiner, P.~A. Kollman, D.~A. Case, U.~C. Singh, C.~Ghio, G.~Alagona,
  S.~Profeta, P.~Weiner, G.~Alagona, P.~Weinerl, {A new force field for
  molecular mechanical simulation of nucleic acids and proteins A New Force
  Field for Molecular Mechanical Simulation of Nucleic Acids and Proteins},
  Journal of the American Chemical Society 106~(3) (1984) 765--784.
\newblock \href {http://dx.doi.org/10.1021/ja00315a051}
  {\path{doi:10.1021/ja00315a051}}.

\bibitem{schutt2017quantum}
K.~T. Sch{\"u}tt, F.~Arbabzadah, S.~Chmiela, K.~R. M{\"u}ller, A.~Tkatchenko,
  Quantum-chemical insights from deep tensor neural networks, Nature
  communications 8 (2017) 13890.

\bibitem{faber2016machine}
F.~A. Faber, A.~Lindmaa, O.~A. von Lilienfeld, R.~Armiento, {Machine Learning
  Energies of 2 Million Elpasolite (A B C 2 D 6) Crystals}, Physical Review
  Letters 117~(13) (2016) 135502.

\bibitem{chmiela2017machine}
S.~Chmiela, A.~Tkatchenko, H.~E. Sauceda, I.~Poltavsky, K.~T. Sch{\"{u}}tt,
  K.-R. M{\"{u}}ller, {Machine learning of accurate energy-conserving molecular
  force fields}, Science Advances 3~(5) (2017) e1603015.

\bibitem{Gomez-Bombarelli2016}
R.~G{\'{o}}mez-Bombarelli, D.~Duvenaud, J.~M. Hern{\'{a}}ndez-Lobato,
  J.~Aguilera-Iparraguirre, T.~D. Hirzel, R.~P. Adams, A.~Aspuru-Guzik,
  {Automatic chemical design using a data-driven continuous representation of
  molecules} (2016) 1--28\href {http://arxiv.org/abs/1610.02415}
  {\path{arXiv:1610.02415}}.

\bibitem{rupp2015machine}
M.~Rupp, R.~Ramakrishnan, O.~A. von Lilienfeld, {Machine learning for quantum
  mechanical properties of atoms in molecules}, The Journal of Physical
  Chemistry Letters 6~(16) (2015) 3309--3313.

\bibitem{ramakrishnan2015big}
R.~Ramakrishnan, P.~O. Dral, M.~Rupp, O.~A. von Lilienfeld, {Big data meets
  quantum chemistry approximations: the $\Delta$-machine learning approach},
  Journal of chemical theory and computation 11~(5) (2015) 2087--2096.

\bibitem{huo2017unified}
H.~Huo, M.~Rupp, {Unified Representation for Machine Learning of Molecules and
  Crystals}, arXiv preprint arXiv:1704.06439.

\bibitem{Artrith2016135}
N.~Artrith, A.~Urban, An implementation of artificial neural-network potentials
  for atomistic materials simulations: Performance for tio2, Computational
  Materials Science 114 (2016) 135 -- 150.
\newblock \href
  {http://dx.doi.org/https://doi.org/10.1016/j.commatsci.2015.11.047}
  {\path{doi:https://doi.org/10.1016/j.commatsci.2015.11.047}}.

\bibitem{artrith2017efficient}
N.~Artrith, A.~Urban, G.~Ceder, Efficient and accurate machine-learning
  interpolation of atomic energies in compositions with many species, arXiv
  preprint arXiv:1706.06293.

\bibitem{LBWB16}
L.~Li, T.~E. Baker, S.~R. White, K.~Burke, Pure density functional for strong
  correlation and the thermodynamic limit from machine learning, Phys. Rev. B
  94 (2016) 245129.
\newblock \href {http://dx.doi.org/10.1103/PhysRevB.94.245129}
  {\path{doi:10.1103/PhysRevB.94.245129}}.

\bibitem{rupp2012fast}
M.~Rupp, A.~Tkatchenko, K.-R. M{\"u}ller, O.~A. Von~Lilienfeld, Fast and
  accurate modeling of molecular atomization energies with machine learning,
  Physical review letters 108~(5) (2012) 058301.

\bibitem{li2016understanding}
L.~Li, J.~C. Snyder, I.~M. Pelaschier, J.~Huang, U.-N. Niranjan, P.~Duncan,
  M.~Rupp, K.-R. M{\"u}ller, K.~Burke, Understanding machine-learned density
  functionals, International Journal of Quantum Chemistry 116~(11) (2016)
  819--833.

\bibitem{bartok2013representing}
A.~P. Bart{\'o}k, R.~Kondor, G.~Cs{\'a}nyi, On representing chemical
  environments, Physical Review B 87~(18) (2013) 184115.

\bibitem{weininger1988smiles}
D.~Weininger, {SMILES, a chemical language and information system. 1.
  Introduction to methodology and encoding rules}, Journal of chemical
  information and computer sciences 28~(1) (1988) 31--36.

\bibitem{behler2007generalized}
J.~Behler, M.~Parrinello, {Generalized neural-network representation of
  high-dimensional potential-energy surfaces}, Physical review letters 98~(14)
  (2007) 146401.

\bibitem{de2016comparing}
S.~De, A.~P. Bart{\'o}k, G.~Cs{\'a}nyi, M.~Ceriotti, Comparing molecules and
  solids across structural and alchemical space, Physical Chemistry Chemical
  Physics 18~(20) (2016) 13754--13769.

\bibitem{Silver2016}
D.~Silver, A.~Huang, C.~J. Maddison, A.~Guez, L.~Sifre, G.~van~den Driessche,
  J.~Schrittwieser, I.~Antonoglou, V.~Panneershelvam, M.~Lanctot, S.~Dieleman,
  D.~Grewe, J.~Nham, N.~Kalchbrenner, I.~Sutskever, T.~Lillicrap, M.~Leach,
  K.~Kavukcuoglu, T.~Graepel, D.~Hassabis, {Mastering the game of Go with deep
  neural networks and tree search}, Nature 529~(7587) (2016) 484--489.
\newblock \href {http://dx.doi.org/10.1038/nature16961}
  {\path{doi:10.1038/nature16961}}.

\bibitem{Mnih2013}
V.~Mnih, K.~Kavukcuoglu, D.~Silver, A.~a. Rusu, J.~Veness, M.~G. Bellemare,
  A.~Graves, M.~Riedmiller, A.~K. Fidjeland, G.~Ostrovski, S.~Petersen,
  C.~Beattie, A.~Sadik, I.~Antonoglou, H.~King, D.~Kumaran, D.~Wierstra,
  S.~Legg, D.~Hassabis, {Human-level control through deep reinforcement
  learning}, Nature 518~(7540) (2015) 529--533.
\newblock \href {http://dx.doi.org/10.1038/nature14236}
  {\path{doi:10.1038/nature14236}}.

\bibitem{mills2017deep}
K.~Mills, M.~Spanner, I.~Tamblyn, {Deep learning and the Schr\"odinger
  equation}, arXiv preprint arXiv:1702.01361\href
  {http://arxiv.org/abs/1702.01361} {\path{arXiv:1702.01361}}.

\bibitem{schutt2017moleculenet}
K.~T. Sch{\"u}tt, P.-J. Kindermans, H.~E. Sauceda, S.~Chmiela, A.~Tkatchenko,
  K.-R. M{\"u}ller, Moleculenet: A continuous-filter convolutional neural
  network for modeling quantum interactions, arXiv preprint arXiv:1706.08566.

\bibitem{Brockherde}
F.~Brockherde, L.~Vogt, L.~Li, M.~E. Tuckerman, K.~Burke, {By-passing the
  Kohn-Sham equations with machine learning} (2017) 1--13\href
  {http://arxiv.org/abs/1609.02815} {\path{arXiv:1609.02815}}.

\bibitem{VASP1}
G.~Kresse, J.~Hafner, {Ab initio molecular dynamics for liquid metals},
  Physical Review B 47~(1) (1993) 558--561.
\newblock \href {http://dx.doi.org/10.1103/PhysRevB.47.558}
  {\path{doi:10.1103/PhysRevB.47.558}}.

\bibitem{VASP2}
G.~Kresse, J.~Hafner, {Ab initio molecular-dynamics simulation of the
  liquid-metal–amorphous-semiconductor transition in germanium}, Physical
  Review B 49~(20) (1994) 14251--14269.
\newblock \href {http://dx.doi.org/10.1103/PhysRevB.49.14251}
  {\path{doi:10.1103/PhysRevB.49.14251}}.

\bibitem{VASP3}
G.~Kresse, J.~Furthm{\"{u}}ller, {Efficiency of ab-initio total energy
  calculations for metals and semiconductors using a plane-wave basis set},
  Computational Materials Science 6~(15) (1996) 15--50.
\newblock \href {http://arxiv.org/abs/0927-0256(96)00008}
  {\path{arXiv:0927-0256(96)00008}}, \href
  {http://dx.doi.org/10.1016/0927-0256(96)00008-0}
  {\path{doi:10.1016/0927-0256(96)00008-0}}.

\bibitem{VASP4}
G.~Kresse, {Efficient iterative schemes for ab initio total-energy calculations
  using a plane-wave basis set}, Physical Review B 54~(16) (1996) 11169--11186.
\newblock \href {http://dx.doi.org/10.1103/PhysRevB.54.11169}
  {\path{doi:10.1103/PhysRevB.54.11169}}.

\bibitem{perdew1996generalized}
J.~P. Perdew, K.~Burke, M.~Ernzerhof, Generalized gradient approximation made
  simple, Physical review letters 77~(18) (1996) 3865.

\bibitem{abadi2016tensorflow}
M.~Abadi, A.~Agarwal, P.~Barham, E.~Brevdo, Z.~Chen, C.~Citro, G.~S. Corrado,
  A.~Davis, J.~Dean, M.~Devin, et~al., Tensorflow: Large-scale machine learning
  on heterogeneous distributed systems, arXiv preprint arXiv:1603.04467.

\bibitem{dewancker2016evaluation}
I.~Dewancker, M.~McCourt, S.~Clark, P.~Hayes, A.~Johnson, G.~Ke, Evaluation
  system for a bayesian optimization service, arXiv preprint arXiv:1605.06170.

\bibitem{kingma2014adam}
D.~Kingma, J.~Ba, Adam: A method for stochastic optimization, arXiv preprint
  arXiv:1412.6980.

\bibitem{pedregosa2011scikit}
F.~Pedregosa, G.~Varoquaux, A.~Gramfort, V.~Michel, B.~Thirion, O.~Grisel,
  M.~Blondel, P.~Prettenhofer, R.~Weiss, V.~Dubourg, et~al., Scikit-learn:
  Machine learning in python, Journal of Machine Learning Research 12~(Oct)
  (2011) 2825--2830.

\end{thebibliography}

\end{document}